\documentclass[chicago,numberedappendix,tighten]{emulateapj}

\usepackage{amsmath,amssymb,natbib,graphicx}
\usepackage{color}
\usepackage{ulem}

\slugcomment{DRAFT \today}

\shorttitle{A Missing-Link in the Supernova-GRB Connection}
\shortauthors{Chakraborti et al.}

\begin{document}

%% LaTeX will automatically break titles if they run longer than
%% one line. However, you may use \\ to force a line break if
%% you desire.

\title{A Missing-Link in the Supernova-GRB Connection: The Case of SN 2012ap}

%% Use \author, \affil, and the \and command to format
%% author and affiliation information.
%% Note that \email has replaced the old \authoremail command
%% from AASTeX v4.0. You can use \email to mark an email address
%% anywhere in the paper, not just in the front matter.
%% As in the title, use \\ to force line breaks.

\author{Sayan Chakraborti\altaffilmark{1,2}, Alicia Soderberg\altaffilmark{1}, Laura Chomiuk\altaffilmark{3}, Atish Kamble\altaffilmark{1}, Naveen Yadav\altaffilmark{4},
Alak Ray\altaffilmark{4}, Kevin Hurley\altaffilmark{5}, Raffaella Margutti\altaffilmark{1}, Dan Milisavljevic\altaffilmark{1},
Michael Bietenholz\altaffilmark{6,7}, Andreas Brunthaler\altaffilmark{8}, Giuliano Pignata\altaffilmark{9}, Elena Pian\altaffilmark{10}, Paolo Mazzali\altaffilmark{11,12},
Claes Fransson\altaffilmark{13}, Norbert Bartel\altaffilmark{7}, Mario Hamuy\altaffilmark{14}, Emily Levesque\altaffilmark{15}, Andrew MacFadyen\altaffilmark{16},
Jason Dittmann\altaffilmark{1}, Miriam Krauss\altaffilmark{17}, M. S. Briggs\altaffilmark{18}, V. Connaughton\altaffilmark{18},K. Yamaoka\altaffilmark{19},
T. Takahashi\altaffilmark{20}, M. Ohno\altaffilmark{21}, Y. Fukazawa\altaffilmark{21}, M. Tashiro\altaffilmark{22}, Y. Terada\altaffilmark{23}, T. Murakami\altaffilmark{23},
J. Goldsten\altaffilmark{24}, S. Barthelmy\altaffilmark{25}, N. Gehrels\altaffilmark{25}, J. Cummings\altaffilmark{25,26}, H. Krimm\altaffilmark{25,27},
D. Palmer\altaffilmark{28}, S. Golenetskii\altaffilmark{29}, R. Aptekar\altaffilmark{29}, D. Frederiks\altaffilmark{29},
D. Svinkin\altaffilmark{29}, T. Cline\altaffilmark{30}, I. G. Mitrofanov\altaffilmark{31}, D. Golovin\altaffilmark{31}, M. L. Litvak\altaffilmark{31}, A. B. Sanin\altaffilmark{31},
W. Boynton\altaffilmark{32}, C. Fellows\altaffilmark{32}, K. Harshman\altaffilmark{32}, H. Enos\altaffilmark{32}, A. von Kienlin\altaffilmark{33},
A. Rau\altaffilmark{33}, X. Zhang\altaffilmark{33}, V. Savchenko\altaffilmark{34}.}

 \altaffiltext{1}{Institute for Theory and Computation, Harvard-Smithsonian Center for Astrophysics, 60 Garden Street, Cambridge, MA 02138, USA}
 \altaffiltext{2}{Society of Fellows, Harvard University, 78 Mt. Auburn Street, Cambridge, MA 02138, USA}
 \altaffiltext{3}{Dept. of Physics and Astronomy, Michigan State University, East Lansing, MI 48824, USA}
 \altaffiltext{4}{Tata Institute of Fundamental Research, 1 Homi Bhabha Road, Mumbai 400005, India}
 \altaffiltext{5}{Space Sciences Laboratory, University of California, 7 Gauss Way, Berkeley, CA 94720, USA}
 \altaffiltext{6}{Department of Physics and Astronomy, York University, 4700 Keele St., M3J 1P3 Ontario, Canada}
 \altaffiltext{7}{Hartebeesthoek Radio Astronomy Observatory, PO Box 443, Krugersdrop, 1740, South Africa}
 \altaffiltext{8}{Max-Planck-Institut f\"ur Radioastronomie, Auf dem H\"ugel 69, 53121 Bonn, Germany}
 \altaffiltext{9}{Departamento de Ciencias Fisicas, Universidad Andres Bello, Avda. Republica 252, Santiago, Chile}
 \altaffiltext{10}{Scuola Normale Superiore, Piazza Dei Cavalieri 7 - 56126 Pisa, Italy}
 \altaffiltext{11}{Liverpool John Moores University, IC2, 146 Brownlow Hill, Liverpool, United Kingdom}
 \altaffiltext{12}{Max-Planck Institute for Astrophysics, Karl-Schwarzschild-Str. 1, 85748 Garching, Germany}
 \altaffiltext{13}{Department of Astronomy, Stockholm University, AlbaNova, SE-106 91 Stockholm, Sweden}
 \altaffiltext{14}{Departamento de Astronomía, Universidad de Chile, Chile}
 \altaffiltext{15}{University of Colorado, C327A, Boulder, CO 80309, USA}
 \altaffiltext{16}{New York University, 4 Washington Place, New York, NY 10003, USA}
 \altaffiltext{17}{National Radio Astronomy Observatory, P.O. Box 0, Socorro, NM 87801, USA}
 \altaffiltext{18}{The Center for Space Plasma and Aeronomic Research, University of Alabama in Huntsville, Huntsville, AL 35899, USA}
 \altaffiltext{19}{Graduate School of Science, Nagoya University, Furo-cho, Nagoya 464-8602, Japan}
 \altaffiltext{20}{ISAS JAXA, 3-1-1 Yoshinodai, Chuo-ku, Sagamihara, Kanagawa 252-5210, Japan}
 \altaffiltext{21}{Hiroshima University, 1-3-1 Kagamiyama, Higashi-Hiroshima, Hiroshima 739-8526, Japan}
 \altaffiltext{22}{Saitama University, 255 Shimo-Okubo, Sakura-ku, Saitama-shi, Saitama 338-8570, Japan}
 \altaffiltext{23}{Kanazawa University, Kadoma-cho. Kanazawa, Ishikawa 920-1192, Japan}
 \altaffiltext{24}{Applied Physics Laboratory, Johns Hopkins University, Laurel, MD 20723, USA}
 \altaffiltext{25}{NASA Goddard Space Flight Center, Greenbelt, MD 20771, USA}
 \altaffiltext{26}{UMBC Physics Department, 1000 Hilltop Circle, Baltimore, MD 21250, USA}
 \altaffiltext{27}{Universities Space Research Association, 10211 Wincopin Circle, Columbia MD 20144, USA}
 \altaffiltext{28}{Los Alamos National Laboratory, Los Alamos, NM 87545, USA}
 \altaffiltext{29}{Ioffe Physical Technical Institute, St. Petersburg, 194021, Russia}
 \altaffiltext{30}{Emeritus, NASA Goddard Space Flight Center, Greenbelt MD 20771, USA}
 \altaffiltext{31}{Space Research Institute, 84/32, Profsoyuznaya, Moscow 117997, Russia}
 \altaffiltext{32}{Department of Planetary Sciences, University of Arizona, Tucson, AZ 85721, USA}
 \altaffiltext{33}{MPE, Giessenbachstrasse, Postfach 1312, D-85748 Garching, Germany}
 \altaffiltext{34}{Observatoire de Paris, 10 rue Alice Domon et Leonie Duquet, F-75205 Paris Cedex 13, France}

\email{schakraborti@fas.harvard.edu}

%% Notice that each of these authors has alternate affiliations, which
%% are identified by the \altaffilmark after each name.  Specify alternate
%% affiliation information with \altaffiltext, with one command per each
%% affiliation.

%% Mark off your abstract in the ``abstract'' environment. In the manuscript
%% style, abstract will output a Received/Accepted line after the
%% title and affiliation information. No date will appear since the author
%% does not have this information. The dates will be filled in by the
%% editorial office after submission.

\begin{abstract}
Gamma Ray Bursts (GRBs) are characterized by ultra-relativistic
outflows, while supernovae are generally characterized by non-relativistic
ejecta.
GRB afterglows decelerate rapidly usually within days,  because their
low-mass ejecta rapidly sweep up a comparatively larger mass of
circumstellar material.
However supernovae, with heavy ejecta, can be in nearly free expansion for centuries.
Supernovae were thought to have non-relativistic outflows except for few
relativistic ones accompanied by GRBs. This clear division was blurred by
SN 2009bb, the first supernova with a relativistic
outflow without an observed GRB. Yet the ejecta from SN 2009bb was baryon
loaded, and in nearly-free expansion for a year, unlike GRBs.
We report the first supernova discovered
without a GRB, but with rapidly decelerating mildly relativistic
ejecta, SN 2012ap. We discovered a bright and rapidly evolving radio counterpart
driven by the circumstellar interaction of the relativistic ejecta. However,
we did not find any coincident GRB with an isotropic fluence of more than
a sixth of the fluence from GRB 980425. This shows for the first time that central engines in
type Ic supernovae, even without an observed GRB, can produce both
relativistic and rapidly decelerating outflows like GRBs.
\end{abstract}

%% Keywords should appear after the \end{abstract} command. The uncommented
%% example has been keyed in ApJ style. See the instructions to authors
%% for the journal to which you are submitting your paper to determine
%% what keyword punctuation is appropriate.

\keywords{gamma rays: bursts --- supernovae: individual (SN 2012ap) --- shock waves
--- radiation mechanisms: non-thermal --- techniques: interferometric}

%% From the front matter, we move on to the body of the paper.
%% In the first two sections, notice the use of the natbib \citep
%% and \citet commands to identify citations.  The citations are
%% tied to the reference list via symbolic KEYs. The KEY corresponds
%% to the KEY in the \bibitem in the reference list below. We have
%% chosen the first three characters of the first author's name plus
%% the last two numeral of the year of publication as our KEY for
%% each reference.

%% Authors who wish to have the most important objects in their paper
%% linked in the electronic edition to a data center may do so by tagging
%% their objects with \objectname{} or \object{}.  Each macro takes the
%% object name as its required argument. The optional, square-bracket 
%% argument should be used in cases where the data center identification
%% differs from what is to be printed in the paper.  The text appearing 
%% in curly braces is what will appear in print in the published paper. 
%% If the object name is recognized by the data centers, it will be linked
%% in the electronic edition to the object data available at the data centers  
%%
%% Note that for sources with brackets in their names, e.g. [WEG2004] 14h-090,
%% the brackets must be escaped with backslashes when used in the first
%% square-bracket argument, for instance, \object[\[WEG2004\] 14h-090]{90}).
%%  Otherwise, LaTeX will issue an error. 

\section{Introduction}
\subsection{Ordinary Supernovae}
The optical lightcurves of supernovae have been well described by \citet{1982ApJ...253..785A} as a
nearly black body photosphere which recedes into the ejecta, heated by $\gamma$-rays
from nuclear decay and cooled by rapid expansion with a characteristic velocity
of $v\sim 10^4$ km s$^{-1}$. The interaction of the ejecta with the circumstellar medium
set up by the stellar wind of the progenitor star has been described by \citet{1982ApJ...258..790C}
using self similar solutions. Such solutions have also successfully described the
radio emission from type Ic supernovae \citep{1998ApJ...499..810C}.
The combination of nearly free expansion leading to decreasing optical thickness
from free-free or synchrotron self absorption (SSA) and decreasing magnetic fields produces
a decreasing peak frequency (the frequency at which the peak in the radio spectrum occurs),
but a nearly constant flux density at the peak frequency \citep{1998ApJ...499..810C}.
In the self-similar solution, this interaction produces a shockfront which
expands in a powerlaw fashion, with $R \propto t^m$, where $R$ is the
radius of the shockfront, $t$ is the time since shock breakout and $m$
is known as the expansion parameter (also sometimes called the deceleration parameter).
In the case of Type Ib/c
supernovae, which generally have a relatively tenuous circumstellar
medium, this interaction produces little deceleration, and $m$ is
close to 1. 
The ejecta would slow down significantly only after encountering a mass of 
external medium comparable to the $\sim1$ M$_\odot$ of ejecta expected for Type I b/c supernovae.
This Sedov time \citep{1950RSPSA.201..159T} is expected to be $\gtrsim 10^2$ years for supernovae.
Therefore young
supernovae are usually found in a Newtonian phase of nearly free expansion.

\subsection{Gamma Ray Bursts}
GRBs were discovered by  \cite{1973ApJ...182L..85K} using the Vela satellites,
designed for the detection of nuclear tests in space.
Ultra-relativistic blast waves had already been described by fluid dynamical
\citet{1976PhFl...19.1130B} solutions before they were implicated in the production
of GRBs \citep{1986ApJ...308L..47G,1986ApJ...308L..43P}. The production of GRBs
require the ejecta to have initial bulk Lorentz factors of $\Gamma \gtrsim 10^2$ \citep{1999PhR...314..575P}
in order to overcome the pair production opacity \citep{1975NYASA.262..164R,1978Natur.271..525S}.
Furthermore, even a small number of baryons in the initial ejecta can soak up most of
the explosion energy available for $\gamma$-rays \citep{1990ApJ...365L..55S}.
This is called baryon poisoning. So GRB jets must have
$\lesssim 10^{-6}$ M$_\odot$ of relativistic ejecta \citep{1999PhR...314..575P}
after breakout from the stellar progenitor surface,
so as not to be baryon poisoned. In a
short time, this relatively small mass of ejecta encounters a
larger mass of external matter. Therefore GRB afterglows are found in
a relativistic but rapidly decelerating phase and have $\Gamma \lesssim 20$. The
broad-lined (high velocity) type Ic SN 1998bw
was discovered in the direction of GRB980425 \citep{1998Natur.395..670G}, and was
characterized by the bright radio emission from its relativistic ejecta 
which had $\Gamma \sim 2-3$ \citep{1998Natur.395..663K}.
This association between a Type Ic supernova and a long GRB was followed by other
supernovae like that of GRB030329 with SN 2003dh \citep{2003Natur.423..847H}
and XRF060218 with SN 2006aj \citep{2006Natur.442.1011P,2006Natur.442.1014S}
also associated with GRBs.
Some of these supernovae were marked by broad lines or by asphericity.
Yet the definitive property, of this small subset of Ic supernovae, that let them produce
relativistic outflows remains elusive.

\subsection{Search for relativistic ejecta}
This evidence, for a supernova-GRB connection, inspired a systematic
search for relativistic ejecta from nearby type Ic supernovae using
radio observations, leading to the discovery of
SN 2009bb by \citet{2010Natur.463..513S} with $\gtrsim10^{49}$ erg of energy in radio emitting
relativistic ejecta.
Like GRBs, SN 2009bb had relativistic
ejecta \citep{2010Natur.463..513S} but this ejecta continued to be in nearly
free expansion for $\sim 1$ year \citep{2011NatCo...2E.175C,2010ApJ...725....4B}, leading to the
suggestion, that unlike GRBs, it is baryon loaded \citep{2011ApJ...729...57C}.
It has been suggested by \citet{2011NatCo...2E.175C} that such engine-driven
relativistic supernovae can even accelerate ultra-high-energy cosmic rays.
Such a baryon loaded fireball has
also been implied in PTF11agg \citep{2013ApJ...769..130C}.
A rapid decline in flux density, faster than $\propto t^{-1}$ at a given radio
frequency, or a decreasing
peak flux density may signal deceleration of the ejecta.
Recently the rapidly declining PTF12gzk was observed with $\lesssim10^{46}$ erg
of energy in fast ejecta \citep{2013ApJ...778...63H}. So far, however, highly energetic
($\gtrsim10^{49}$ erg) relativistic ejecta combined with rapid deceleration has never
been observed until now in a supernova unassociated with a GRB.

\section{Discovery of SN 2012ap}
SN 2012ap was discovered in NGC 1729 by the Lick Observatory Supernova
Search \citep{2012CBET.3037....1J} on 2012 February 10.
NGC 1729 is at a distance of  $D\sim40$ Mpc \citep{2007ApJS..172..599S},
and we will take a distance of $40.0$ Mpc in what follows, although we
note the dependence on distance where appropriate.
The host galaxy was observed on 2012 February 5 and the supernova was not yet
detected on that date. Therefore the explosion must have occurred either after
or at most a few days before this date.
Spectroscopically it was identiﬁed as a type Ic supernova, with broad
lines, and marked by similarities with SN 2009bb and the
GRB-associated SN 1998bw \citep{2012CBET.3037....2M}. \citet{2014ApJ...782L...5M}
reported unusually strong diffuse interstellar bands in
its optical spectra. \citet{2014arXiv1402.6344M} report on the non-detection
of X-ray emission and therefore indicate a shortlived central engine.
\citet{2014arXiv1408.1606M} models the optical emission from the supernova
and infer a explosion date of February 5 with an uncertainty of 2 days.

\begin{table*}
\begin{center}
\caption{Radius-Magnetic Field Evolution\label{table}}
 \begin{tabular}{| l | r r r r r |}
\hline			
Observation	& Age   	& $F_{\nu p}$\tablenotemark{a}	& $\nu_{p}$	& $R$   	& $B$\\
Date (2012)   	& (Days)	& (mJy)   	& (GHz)   	& ($10^{15}$cm)	& (mG)\\
\hline
February 17.0 	& 12.0	& 5.85$\pm$0.58	& 11.82$\pm$0.48	& 10.7$\pm$0.7	& 1084$\pm$46\\
February 23.2 	& 18.2	& 5.69$\pm$0.07	& 8.92$\pm$0.08	& 14.0$\pm$0.2	& 820$\pm$8\\
March 02.0 	& 27.0	& 4.71$\pm$0.10	& 6.13$\pm$0.09	& 18.7$\pm$0.3	& 576$\pm$8\\
March 12.9 	& 37.9	& 4.20$\pm$0.10	& 4.54$\pm$0.10	& 23.9$\pm$0.6	& 431$\pm$9\\
\hline  
\end{tabular}
\end{center}
\tablenotetext{1}{
Peak flux densities and peak frequencies of SN 2012ap are determined
from our VLA and GMRT observations by fitting a SSA spectrum to the
observed flux densities.
Radius and magnetic fields are determined 
assuming equipartition (a possible source of systematic error). Quoted errors
are statistical $1\sigma$ uncertainties.}
\end{table*}

\begin{figure}
\includegraphics[width=\columnwidth]{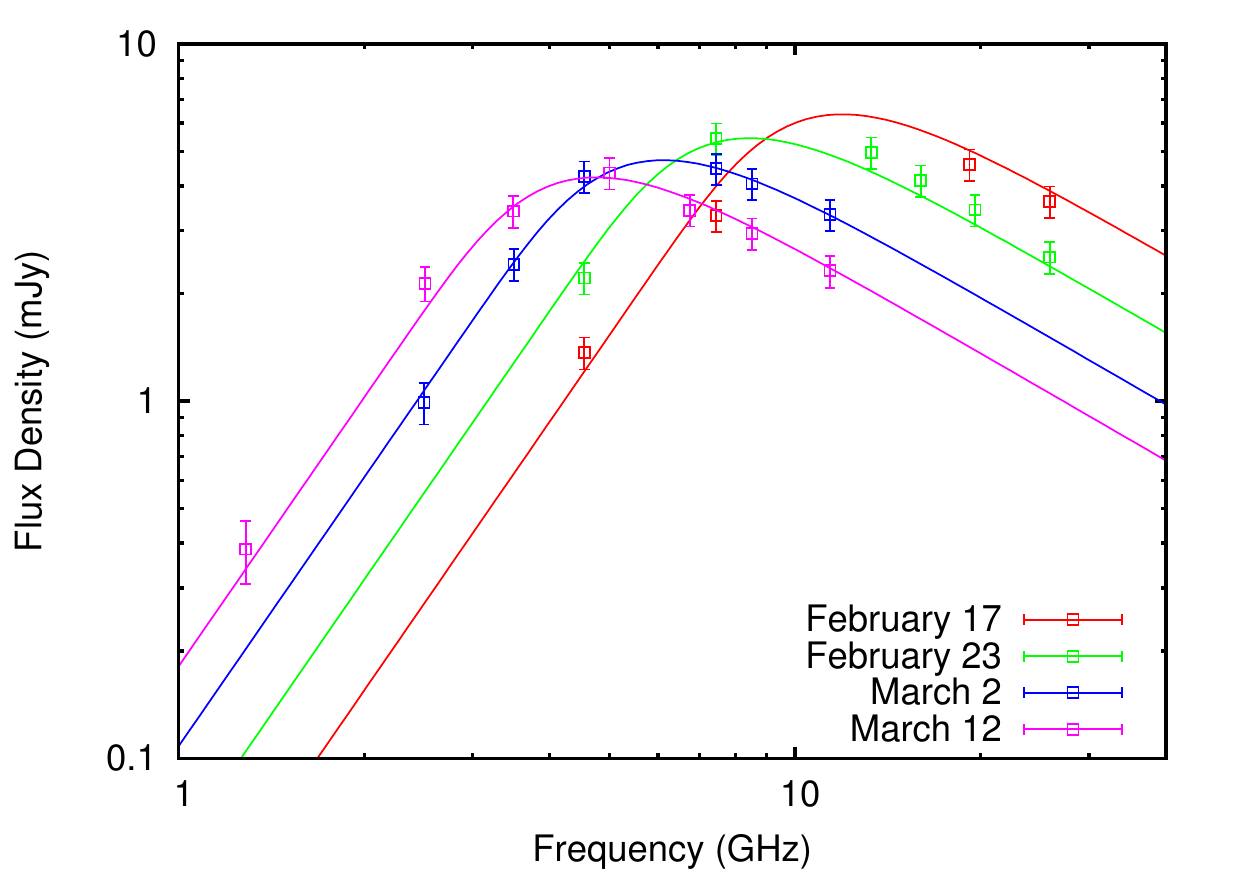}
\caption{\textbf{Temporal and spectral flux density evolution:}
Radio observations of SN 2012ap were obtained using the VLA and GMRT. A simple
synchrotron spectrum, with a low frequency turnover due to
SSA \citep{1998ApJ...499..810C} can explain the emission at all epochs
discussed here. 
Note that the peak frequency decreases with time, like that of
ordinary radio supernovae. In contrast to what is observed in ordinary supernovae
however, the peak flux density also decreases with time. \label{F_nu_t}
}
\end{figure}

We detected strong radio
emission from the supernova using the Karl G. Jansky Very Large Array (VLA) on 2012
February 15.0 UT, with a steep inverted spectrum between 5.0 GHz and 6.75 GHz. The
subsequent radio observations using the VLA and the Giant Metrewave Radio Telescope
(GMRT) showed an SSA spectrum (see Figure \ref{F_nu_t}). See Appendix
for details of radio observations and modeling. The peak flux densities and peak
frequencies are summarized in Table \ref{table}.

\section{Observations}
\begin{figure}
\includegraphics[width=\columnwidth]{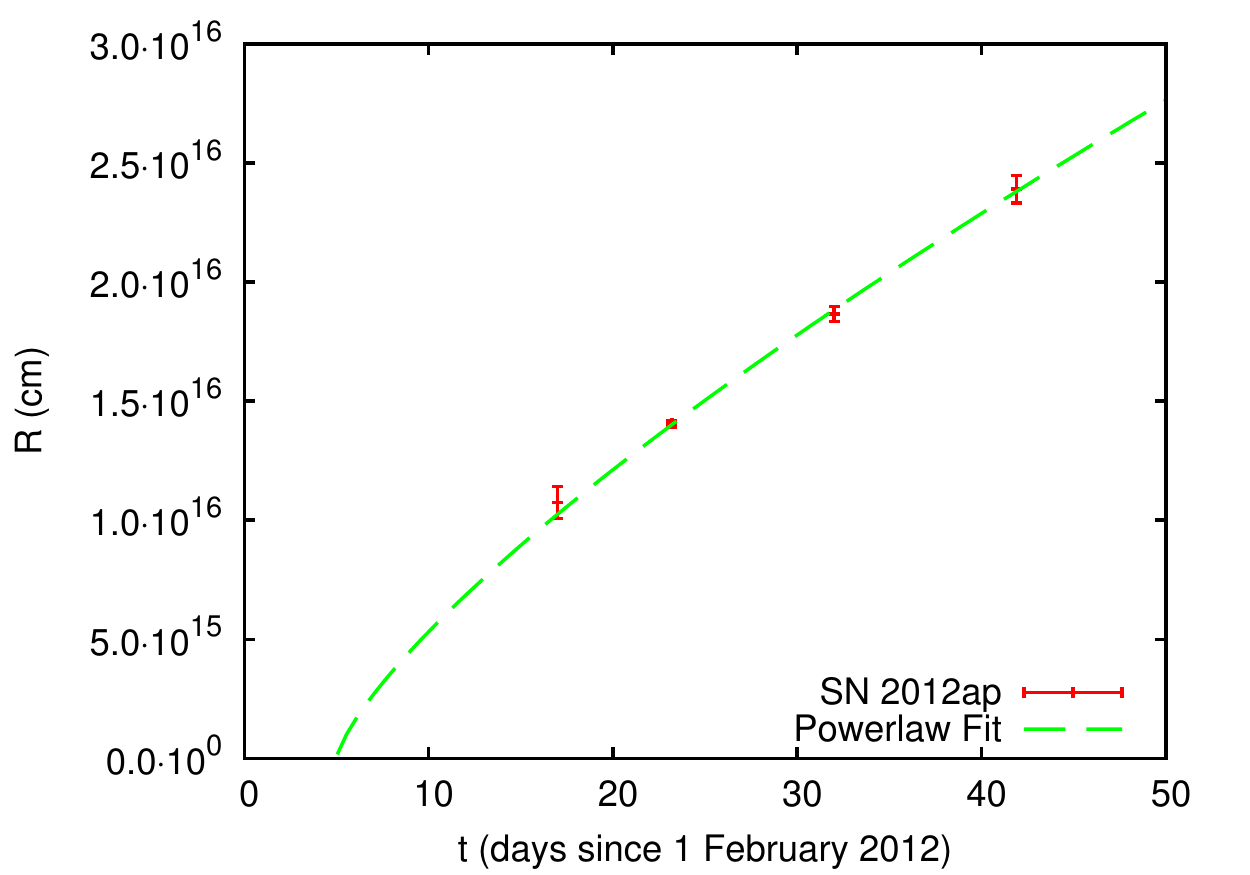}
\caption{\textbf{Temporal size evolution:}
Radii were determined by fitting an SSA spectrum \citep{1998ApJ...499..810C}
to the flux densities at each individual epoch. We assumed an explosion date
of 2012 February 5 for this fit\label{Rt}
}
\end{figure}

In this section we describe the observations of SN 2012ap obtained at various wavelengths.
The first broadband radio spectrum was obtained at an age of 12 days. The radio data was well described with
a SSA spectrum (see Figure \ref{F_nu_t}). This allowed us to estimate
the radius of the emission region \citep{1998ApJ...499..810C} to be $\sim 10^{16}$ cm.
The inferred size evolution as a funtion of time is shown in Figure \ref{Rt}
The expanding ejecta may have slowed down since the time of explosion, therefore this
size gives us a lower limit on the initial
apparent expansion velocity of $0.4c$. This is much faster than all other $>100$ nearby
type Ibc supernovae in our sample \citep{2010Natur.463..513S}, apart from SN 2009bb.
Hence, SN 2012ap was chosen for detailed
follow-up as the second relativistic supernova ever found without an associated GRB.
%Lack of high energy counterpart
This prompted a search for a high energy counterpart using data from all the
spacecraft within the InterPlanetary Network (IPN), a group of spacecraft equipped with gamma-ray
detectors for localizing GRBs \citep{2009AIPC.1133...55H}.
The locations of the transients are determined by comparing the arrival
times of the $\gamma$-rays at the different spacecrafts.
In the plausible range of explosion dates, 2012 February 5 to 10, there
is no evidence for a GRB associated with SN 2012ap down to the IPN threshold
of $6 \times 10^{-7}$ erg cm$^{-2}$. We looked for transients with even
lower fluences using Fermi and Swift, but we found none having small error
boxes consistent with the position of the supernova.

\subsection{Radio Temporal Evolution}
No supernova was seen in the host galaxy on a pre-explosion image
taken by Katzman Automatic Imaging Telescope (KAIT) on February 5
down to an upper limit of 18.7 R magnitude.
A subsequent image taken on February 10, revealed SN 2012ap at 17.3
R magnitude. We therefore conclude that the explosion happened
before February 10, but not much before February 5. Since one of the
major claims of this work, is
a deceleration, we count the age of the supernova from the February
5, which is near the beginning of this range. Later dates would imply
even more deceleration. \citet{2014arXiv1408.1606M} models
the optical lightcurve and provides an explosion date of February
$5 \pm 2$, which is used in this work and interpreted as a $1\sigma$ bound
on the explosion date. The present uncertainties of
a few days in the explosion date dominates the uncertainty in $m$.

Both the VLA and GMRT observations have been reduced using Astronomical Image Processing 
Software ({\it AIPS}) standard techniques.
Radio Frequency Interference in data was flagged and the interferometric
visibilities were amplitude and phase calibrated. Bandpass calibration
was done using BPASS based on the flux calibrators 3C48 and/or 3C147.
The single source data was extracted using AIPS task
SPLIT after final bandpass and flux calibration. The single
source data sets were imaged using IMAGR. The images were
corrected for the residual phase calibration errors using self-calibration
of visibility phases \citep{1989ASPC....6..185C}.
The source flux densities were extracted
by fitting Gaussian using task JMFIT assuming point sources.
The errors reported on the flux density are obtained by using the image
statistics from the region surrounding the source.

\begin{figure}
\includegraphics[width=\columnwidth]{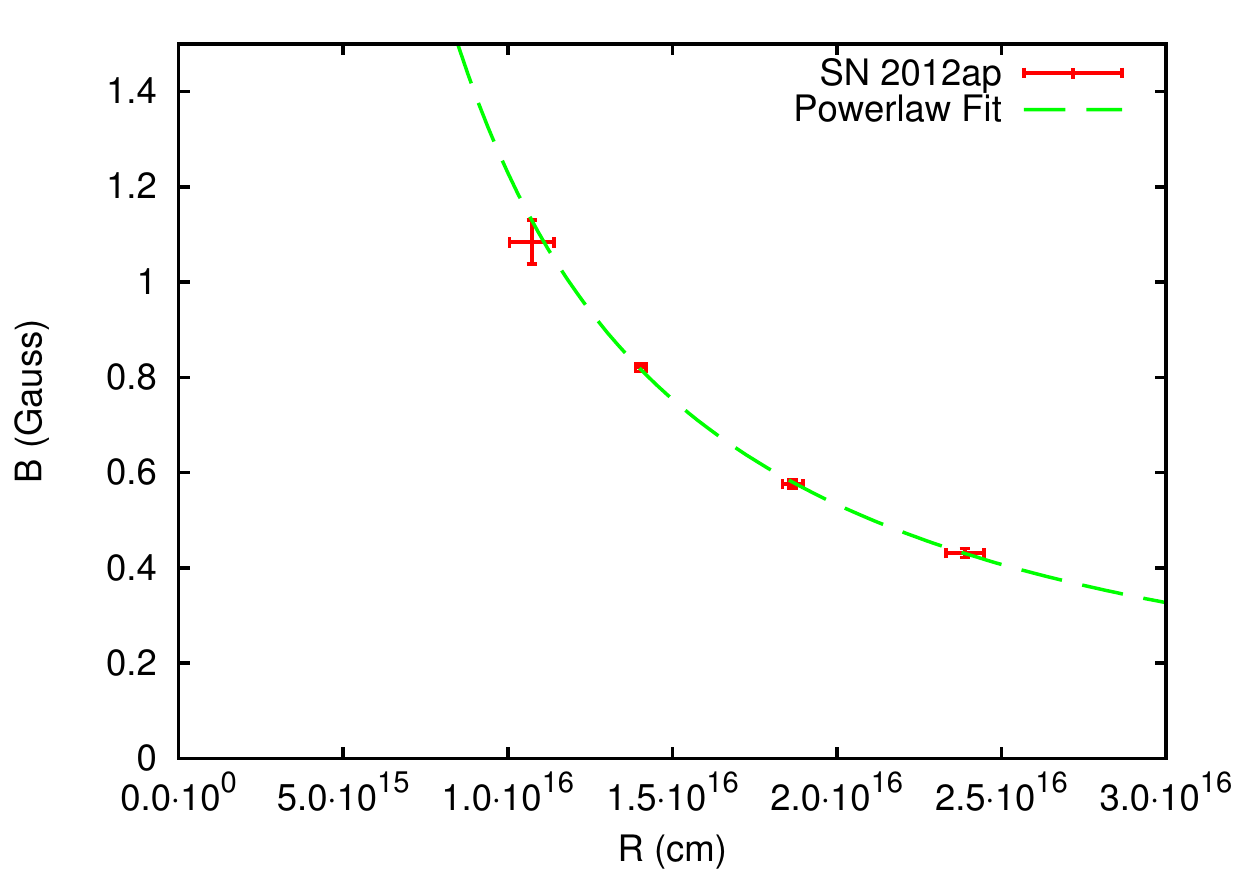}
\caption{\textbf{Magnetic field evolution:}
Radii and magnetic fields were determined by fitting an SSA
spectrum \citep{1998ApJ...499..810C} to the flux densities at each
individual epoch.\label{BR}
}
\end{figure}

These radio observations were used to obtain the values reported in Table \ref{table}
and in Figures \ref{Rt} and \ref{BR}.
Quoted uncertainties in flux densities are statistical. There may be up to $5\%$ systematic
uncertainty from flux density calibration. This dominates the error budget
quoted in the final derived values for $E_0$ and $\dot{M}$.

The full set of radio observations were fit with an SSA model 
with $p=3$, as has been observed in Ic supernovae. Both
the peak flux density and peak frequency were allowed to vary as arbitrary powerlaws
with an origin at 6th February. A good global fit is obtained with a
$\chi^2=0.98$ per degree of freedom. See Figure \ref{F_nu_t} for all
the radio observations and fit. The radii (Figure \ref{Rt}) and magnetic
fields (Figure \ref{BR}) reported in Table \ref{table} are derived from
fits to the data at individual epochs.

The dependence of $R$ on the ratio
$\epsilon_e / \epsilon_B$ and $f$ are through a power of $1/19$, so they are
likely to introduce only a negligible systematic uncertainty \citep{1998ApJ...499..810C}.
The dependence on distance is with a power $18/19$ which could introduce a
systematic shift, but not change the expansion parameter. The dependence
on the flux density is a power $9/19$, so a $5\%$ error in flux density will
add $2.5\%$ of systematic error to the radii. These errors should be considered
above and beyond the statistical uncertainties reported in the table. However,
none of them are big enough to significantly change the results reported in this work.

\subsection{Search for high energy counterpart}
Between 2012 February 05 and 2012 February 10, inclusive of both days,
a total of 5
bursts were detected by one or more of the nine spacecraft of the
InterPlanetary network (IPN:  Mars Odyssey, Konus-Wind, RHESSI,
INTEGRAL (SPI-ACS), Swift-BAT, Suzaku, AGILE, MESSENGER, and Fermi
(GBM)). All these confirmed bursts were observed by more than one
instrument on one
or more spacecraft, and could be localized at least coarsely.
During
the same period, there were also 3 unconfirmed bursts which were
observed by one experiment on one spacecraft (the 20-detector Suzaku
HXD-WAM) in the triggered (TRN) mode. Their origin is uncertain;
they could be cosmic or solar, but too weak to be detected by other IPN
spacecraft. In at least one case, it could be particle-induced.  They cannot
be localized accurately, but analysis of the detector responses
indicates that they are unlikely to have originated from the direction
of SN 2012ap. Therefore they were excluded from further analysis.  No bursts
from known sources such as AXPs and SGRs were recorded during this
period.

The completeness of our sample can be estimated as follows.  We have 3
distinct sets of events: IPN bursts, Fermi GBM-only bursts, and bursts
observed by the Swift BAT only within its coded field of view. The IPN
is sensitive to bursts with fluences down to about $6 \times 10^{-7}$ erg
cm$^{-2}$, and observes
the entire sky with a temporal duty cycle close to 100\% \citep{2009AIPC.1133...55H}.
This places a threshold of $\sim10^{47}$ erg on the isotropic equivalent $\gamma$-ray
energy of any possibly GRB accompanying SN 2012ap if it were to be detected
by the IPN. This is a factor $\sim 6$ lower than that of GRB980425 accompanying SN 1998bw.

The Fermi GBM detects bursts down to an 8 to 1000 keV fluence of about $4 \times 10^{-8}$
erg cm$^{-2}$, and observes the entire unocculted sky (~8.8 sr) with a
temporal duty cycle of more about 86\%.  The weakest burst observed by
the BAT within its coded field of view had a 15 to 150 keV fluence of
$6 \times 10^{-9}$ erg cm$^{-2}$, and the BAT observes a field of view of about 2 sr
with a temporal duty cycle of about 90\%.  Generally, the
weakest bursts are short-duration GRBs; higher fluences characterize
the sensitivities to long-duration GRBs.

The localization accuracies of the 5 confirmed bursts varied widely.
None were observed within the coded fields of view of the Swift BAT,
INTEGRAL IBIS, or Super-AGILE (several arc minute accuracy).  None were
observed by MAXI (several degree accuracy).
Two were observed either by the Fermi GBM alone, or by the Fermi GBM
and one near-Earth spacecraft (so that they could not be localized
accurately by triangulation).  These bursts had 1 $\sigma$
statistical-only error radii of 23.8 and 5.5 degrees.   The GBM error
contours are not circles, although they are characterized as such, and
have about 3.2 degrees of systematic uncertainties associated with
them.  Adding the statistical and systematic uncertainties in
quadrature gives a reasonable approximation to a 1 $\sigma$ error
circle, and multiplying that radius by 3 gives a reasonable
approximation to a 3 $\sigma$ (statistical and systematic) error circle.  These two
Fermi bursts have positions which are inconsistent with that of the SN
(that is, the SN falls far outside the 3 $\sigma$ error circle as
approximated above).
Two were observed by interplanetary and near-Earth spacecraft,
and could be triangulated to small error boxes whose positions
exclude the position of SN2012ap.
One burst was observed by the Swift BAT outside the coded
field of view, and by the INTEGRAL SPI-ACS.  As the distance
between these two spacecraft is only about 0.5 light-seconds,
the event can be triangulated, but not accurately.  Its error
annulus has an area of 1.8 sr, and it includes the position
of the supernova.
The total area of the localizations of the 5 confirmed bursts was ~0.5
times 4 pi sr.  This implies that about 0.5 bursts can be expected to
have positions which are consistent with any given point on the sky
simply by chance (i.e. within the 3 $\sigma$ error region).  In our sample
one burst has a position consistent with the SN position (Poisson
probability ~0.3).

There is another approach to the probability calculation.  Since only 0
or 1 GRB in our sample can be physically associated with the supernova, we can
calculate two other probabilities.  The first is the probability that,
in our ensemble of 5 bursts, none is associated by chance with the
supernova. For our ensemble, this probability is $0.54$.
The second is that any one burst is associated by chance with the SN,
and that all the others are not. For our ensemble, that probability is $0.39$.
Since we find about the expected number of chance
coincidences, and since there are no bursts with small error boxes
whose positions are consistent with the supernova, there is no strong
evidence for a SN-associated GRB within the time window down to the IPN
threshold.  If the supernova produced a burst below the IPN threshold
and above the Fermi one, it is possible that both Swift and Fermi would
not detect it; considering their spatial and temporal coverages, the
joint non-detection probability is about 0.38.  Finally, if the
supernova produced a burst below the Fermi threshold but above the
Swift one, the probability is about 0.86 that it would not be
detected.

In summary, no confirmed coincident burst with an isotropic 
fluence of more than a sixth of GRB980425 was localized to the direction
of SN 2012ap within the relevant time window. Weaker bursts were searched
for, albeit with $< 100\%$ duty cycle, but also not found.

\section{Interpretation}
In this section we describe what conclusions can be drawn about the explosion,
its energy budget and its environments based on the observations described above.
\subsection{Deceleration}
As a supernova expands, its radio emitting region becomes optically thinner. The expansion
also dilutes the energy density of relativistic electrons and magnetic fields. In the
nearly free expansion phase, these two effects
conspire \citep{1998ApJ...499..810C,2011ApJ...729...57C} to decrease the turnover
frequency of the SSA spectrum but keep the peak flux density nearly constant in time.
In the case of SN 2012ap, the peak flux density fell steadily over the first month
of observations (see Figure \ref{F_nu_t}) and
the expansion is well fit by a power law of the form, $R\propto t ^m$, where $m=0.74\pm0.03$.
Here the 1 $\sigma$ statistical uncertainty is derived from the radio size determinations following
\citet{1998ApJ...499..810C}.
However, the determination of the explosion date derived from optical observations
\citep{2014arXiv1408.1606M} presents a greater 1 $\sigma$ systematic uncertainty of $0.07$.
This is determined by refitting the data with the range of explosion dates allowed
by the optical studies of the supernova.
Taking both into account and summing them in quadrature, the total uncertainly in
$m$ is $0.08$.
Therefore our estimate of the expansion parameter is $m=0.74 \pm 0.08$, which
incorporates our understanding of the statistical and systematic uncertainties
dominated by the radio observations and time of explosion respectively. This
is 3 $\sigma$ away from a value near $1$, expected for undecelerated expansion.

\begin{figure}
\includegraphics[width=\columnwidth]{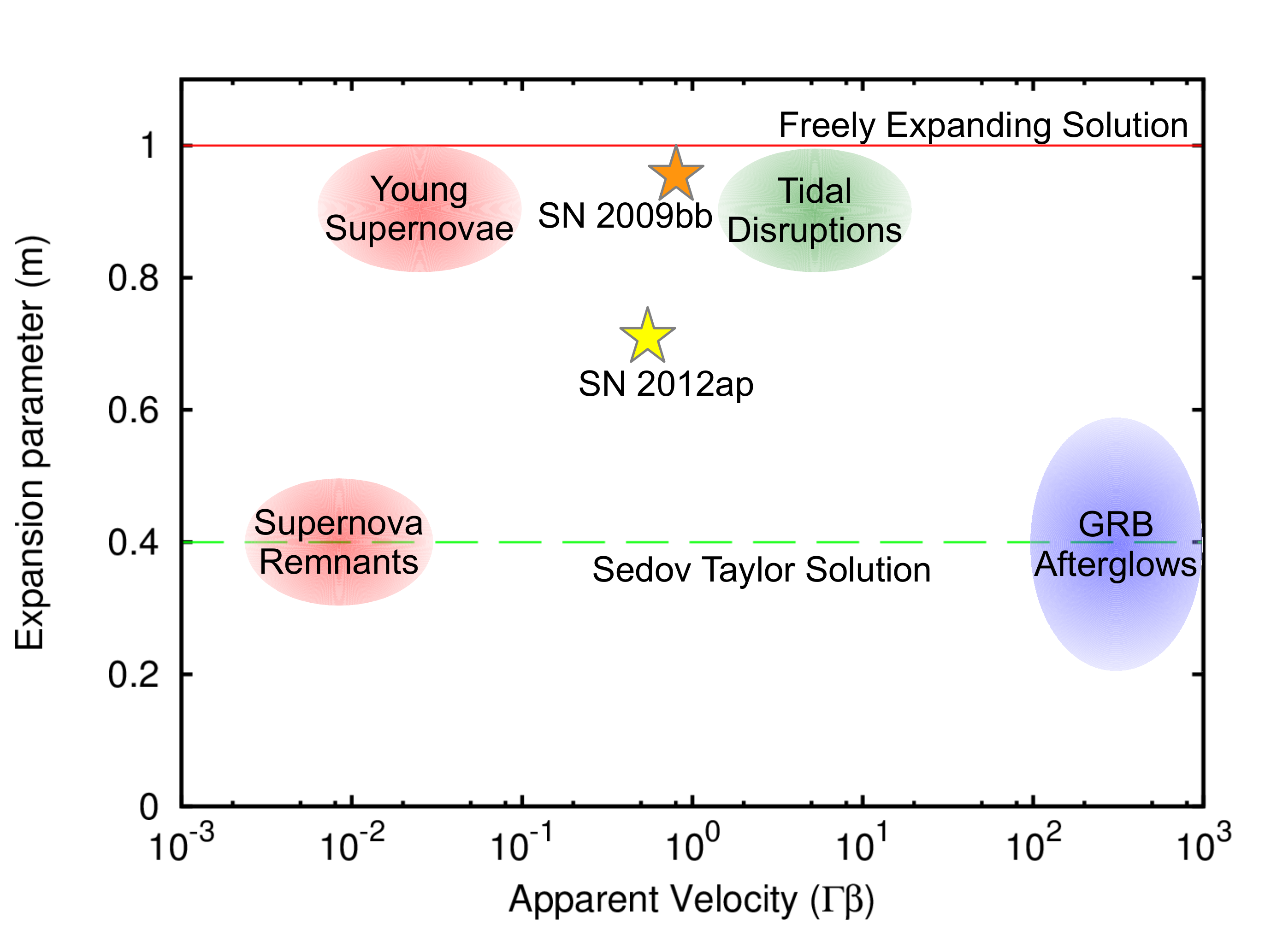}
\caption{\textbf{Phase space of afterglows:}
Explosions are classified in this cartoon diagram by the apparent expansion
velocity $\Gamma\beta$ in natural
units and dimensionless expansion parameter $m$ (where $R\propto t^m$)
of the component driving the radio afterglow. Explosion parameters on the red-line
correspond to freely expanding solutions in which the expansion velocity is constant
in time. Explosions on the green line are in a Sedov Taylor phase \citep{1950RSPSA.201..159T}.
GRB afterglows are found in the lower right portion corresponding to a ultra
relativistic decelerating \citet{1976PhFl...19.1130B} phase.
We indicate the location of supernovae and novae with standard parameters.
SN 2009bb was discovered to be mildly relativistic and in nearly free expansion.
SN 2012ap parameters are determined from radio observations in this work.
The sizes of the blobs represent the typical variation within that class
and the sizes of the stars represent their individual uncertainties.
Note that SN 2012ap populates a hitherto unpopulated phase space.\label{vm_schem}
%(how many Words)
}
\end{figure}

The combination of high velocity and small expansion parameter, place SN 2012ap
at an unique position on the Figure \ref{vm_schem}
Within the context of an ordinary type Ic supernova, this would imply a supernova ejecta
density profile characterized by a powerlaw distribution of mass ejected at a 
particular velocity is $\propto v^{-6}$,
where $v$ is the velocity. However, the expulsion
of stellar envelopes are expected to produce much steeper ejecta
profiles \citep{1999ApJ...510..379M}. This could plausibly be explained by an extraordinarily
energetic supernova with a low ejecta mass and a shallow ejecta profile extending
into relativistic velocities. Our observations of SN 2012ap
imply much more energy coupled to high velocity ejecta than is expected from
realistic explosion models of ordinary core collapse supernovae.
The decelerating expansion
is however, consistent with a Central Engine Driven EXplosion \citep{2011ApJ...729...57C}
(CEDEX) which put a large
amount of energy into a small mass of relativistic ejecta responsible for the radio
afterglow which separates itself from the non relativistic ejecta responsible for the
optical emission. The decelerating expansion of this component interacting with the
wind of the progenitor star is expected to produce (See Appendix)
a $R\propto t^{3/4}$ behavior consistent with the observations.

\subsection{Total energy budget}
\begin{figure}
\includegraphics[width=\columnwidth]{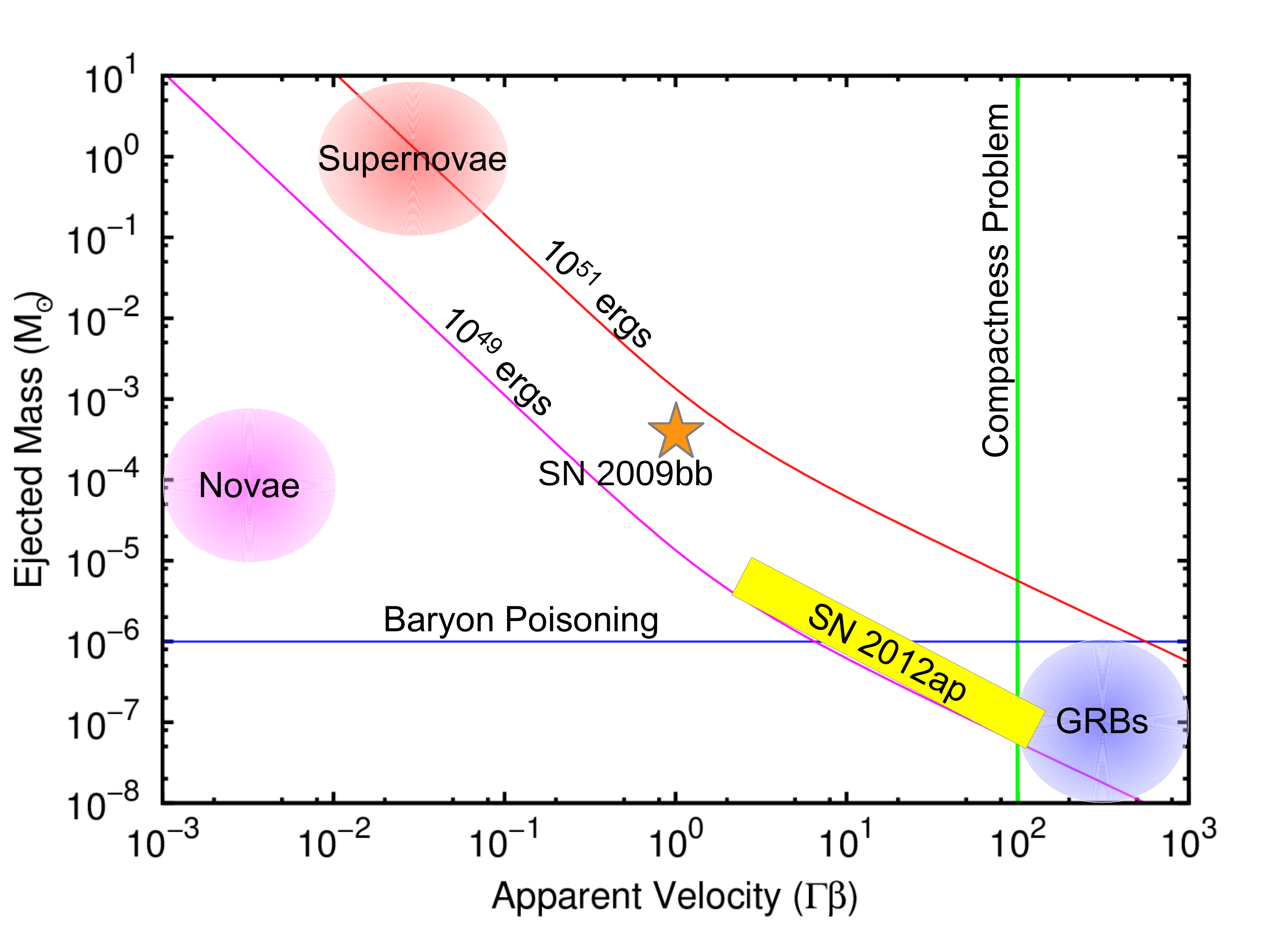}
\caption{\textbf{Phase space of explosions:}
Explosions are classified in this cartoon diagram by their apparent expansion velocity and ejecta mass
responsible for their electromagnetic display. Everything to the left of the green dashed
line suffers from the "compactness problem" \citep{1975NYASA.262..164R,1978Natur.271..525S}
and no $\gamma$-rays can get out due to pair production opacity.
Everything above the blue line is "baryon poisoned" \citep{1990ApJ...365L..55S}
and cannot produce $\gamma$-rays as the baryons take up much of the energy.
So, GRBs are confined to the lower right corner.
We indicate the location of supernovae and novae with standard parameters.
SN 2009bb was discovered to be mildly relativistic
and baryon loaded. SN 2012ap parameters are from blastwave calorimetry in this work.
The red line indicates an explosion energy of $10^{51}$ erg and magenta
line indicates $10^{49}$ erg.
Note that SN 2012ap occupies a hitherto unpopulated region of phase
space in the continuum between supernovae and GRBs, in the region
between SN 2009bb and GRBs.
\label{momemass}
%(how many Words)
}
\end{figure}

Within the context of such a CEDEX, our radio observations allow us to perform
calorimetry of the fireball responsible for this radio afterglow. The initial
energy in the relativistic component can be estimated as (See Appendix
for derivation and assumptions),
\begin{equation}
 E_0=2.35 \times 10^{45} \left(\frac{F_{\nu p}}{{\rm mJy}}\right)^{23/19}
 \left(\frac{\nu_p}{{\rm GHz}}\right)^{-1} \left(\frac{D}{{\rm Mpc}}\right)^{46/19} {\rm erg}
\end{equation}
Using the SSA spectral fits from age 18 days, the epoch with the smallest fractional
uncertainties in
peak flux density $F_{\nu p}$ and peak frequency $\nu_p$, we estimate $(1.6\pm0.1)\times10^{49}$ erg
of energy in the relativistic ejecta. This estimate is robust, irrespective of
whether one chooses a relativistic or Newtonian blastwave model.
This energy is comparable to the energy observed
in the relativistic outflows from SN 2009bb and supernovae such as SN 1998bw
associated with sub-energetic GRBs in the local Universe.
This energy estimate forces the plausible combinations of ejected mass and
initial velocity (see Figure \ref{momemass}) to lie in a narrow region above of
and parallel to the magenta line for $10^{49}$ erg. 

\subsection{Circumstellar density}
In addition, the model allows us to derive (See Appendix) the mass loss
rate of the progenitor, responsible for the circumstellar interaction.
Again using the data from age 18 days, we estimate the pre-explosion mass loss rate 
between $3.5\times10^{-6} {\rm M_\odot yr^{-1}}$ and $4.9\times10^{-5} {\rm M_\odot yr^{-1}}$,
depending on whether one uses a relativistic or Newtonian blastwave model respectively.
This range of values is consistent with the
expected mass loss rate of Wolf Rayet stars \citep{2000A&A...360..227N} and is comparable
to the case of SN 2009bb \citep{2010Natur.463..513S}.
The upper limit on the mass loss rate, derived by \citet{2014arXiv1402.6344M} using X-ray
non detection from the Chandra X-ray observations, rules out the higher mass loss rate
derived from the non-relativistic model.
This encourages us to prefer the lower one of these values.
Hence, we prefer the relativistic blaswave model.
%Ejecta Mass
Since we find the radio afterglow of SN 2012ap in a decelerating phase, its relativistic
outflow must have already swept up a mass of circumstellar matter more than its own mass. This can
put an upper limit on the mass of the relativistic ejecta, which is easily estimated since
we have already determined the circumstellar density and blastwave radius. This argument 
constrains the mass of the relativistic ejecta to be $\lesssim1.2\times10^{-5} {\rm M_\odot}$.
This restricts the allowed range of initial parameters for SN 2012ap (see yellow box in
Figure \ref{momemass}).
Hence, the mass of the ejecta component powering the radio afterglow of SN 2012ap is much
less than that responsible for the optical lightcurve of a typical supernova and closer to
the estimated mass of the relativistic component of a GRB's ejecta.

\section{Discussion}
Even though no GRB counterpart was found for SN 2012ap, its
radio afterglow shares remarkable characteristics with those of GRB associated supernovae; it has a
relativistic ejecta component with relatively few baryons. While SN 2009bb had a relativistic
outflow, it was clearly baryon loaded which is not the case here. A rapid radio
decline has been observed in PTF12gzk which may also have had fast ejecta. But SN 2012ap
had orders of magnitude more energy in relativistic ejecta, putting its energetics firmly
in the class of GRB-associated supernovae. SN 2012ap pushes the boundaries of explosion
parameters observed from stripped core progenitors with central engines
(see Figures \ref{vm_schem} and \ref{momemass}).
The parameters that separate the outflows, from GRB associated supernovae and ordinary
Type Ib/c supernovae, according to \citet{2011ApJ...729...57C} are the velocity and baryon
loading of their fastest ejecta.
According to \citet{2014arXiv1402.6344M} the duration of central engine activity drives
the diversity of explosion outcomes.
By bridging the gap between ordinary supernovae and GRB associated supernovae
in terms of its high velocity and low ejecta mass, SN 2012ap demonstrates the
role of CEDEXs in understanding the supernova-GRB connection.

%% If you wish to include an acknowledgments section in your paper,
%% separate it off from the body of the text using the \acknowledgments
%% command.

%% Included in this acknowledgments section are examples of the
%% AASTeX hypertext markup commands. Use \url without the optional [HREF]
%% argument when you want to print the url directly in the text. Otherwise,
%% use either \url or \anchor, with the HREF as the first argument and the
%% text to be printed in the second.

\acknowledgments

This work made use of radio observations from the NRAO VLA and GMRT. The National Radio
Astronomy Observatory is a facility of the National Science Foundation operated
under cooperative agreement by Associated Universities, Inc. We thank the staff of
the GMRT that made these observations possible. GMRT is run by the National Centre
for Radio Astrophysics of the Tata Institute of Fundamental Research.
The Konus-Wind experiment as part of IPN supported by contract of the Russian Space
Agency and partially supported by the Russian Foundation for Basic Research (grants
12-02-00032 and 13-02-12017). We thank Roger Chevalier and Ehud Nakar for comments.

%\clearpage

%\begin{figure}
%%\epsscale{1.1}
%%\plotone{B_R.eps}
%\includegraphics[width=0.9\columnwidth]{R_t.eps}
%\caption{Evolution of the blast wave radius $R_{lat}$, determined from SSA fit
%to observed radio spectrum, as a function of the time $t_{obs}$ in the observer's frame.
%The evolution is consistent with nearly free expansion. Note that the observations
%require $M_0\gtrsim10^{-2.5}M_\odot$.\label{R_t}}
%\end{figure}

%\newpage
%\clearpage

\appendix

\section{Blastwave Evolution}
We follow the derivation of the blastwave evolution derived for a CEDEX \citep{2011ApJ...729...57C}.
We consider a simple collisional model \citep{1999PhR...314..575P,1999ApJ...512..699C}
where the relativistic ejecta interacts with the circumstellar matter and forms a
decelerating shell. The ejecta is characterized by the
rest frame mass $M_0$ of the shell launched by a CEDEX and its initial Lorentz
factor $\gamma_0$. The total energy is given by $E_0 = \gamma_0 M_0 c^2$. For a
progenitor with mass loss rate of $\dot{M}$ through a wind with velocity $v_w$,
the ejecta sweeps up $AR$ (where $A\equiv \dot{M} / v_w$ and the circumstellar density
falls off as $\propto R^{-2}$) amount of circumstellar
matter within a radius $R$. This slows down the ejecta whose Lorentz factor
evolves as,
\footnotesize
\begin{align}\label{Gt}
&\gamma=\left.\left(\gamma_0 M_0+\frac{-M_0+2 A c \gamma_0 t+\sqrt{8 A c M_0 t \gamma_0^3+(M_0-2 A c \gamma_0 t)^2}}{2 \gamma_0}\right)\right/ \\ \nonumber
&\sqrt{\left(2 A c \gamma_0 t+\sqrt{8 A c M_0 t \gamma_0^3+(M_0-2 A c \gamma_0 t)^2}\right)
   M_0+\frac{\left(-M_0+2 A c \gamma_0 t+\sqrt{8 A c M_0 t \gamma_0^3+(M_0-2 A c \gamma_0 t)^2}\right)^2}{4 \gamma_0^2}},
\end{align}
\normalsize
according to \citet{2011ApJ...729...57C}.

For both the prototypical SN 2009bb and the newly discovered SN 2012ap, the blastwave
was not ultra-relativistic at the time of the observed radio afterglow. Without a large
relativistic beaming effect, the observer would receive emission from the entire
shell of apparent lateral extent $R_{lat}$ at a time $t_{obs}$ given by
\begin{equation}
dt_{obs} = \frac{dR_{lat}}{\beta \gamma c} .
\end{equation}
Radio observations of SN 2012ap measure essentially this transverse $R_{lat}$, not
the line of sight $R$. Substituting $\beta \gamma$ using Equation \ref{Gt} and integrating
gives us a large and unintuitive expression for $R_{lat}(t)$ which is not quoted here.
Since we find
SN 2012ap in a decelerating phase, we take the limit of $M_0 \to 0$ holding $E_0$ constant.
This brings us to the portion of the full solution where the swept up
mass is more than the ejecta mass. Furthermore, since the explosion is still relativistic,
we know that we are still in the early phase and therefore expand the solution as a power
series around $t=0$. This gives us the intermediate asymptotic solution for the lateral expansion
of the blastwave in its decelerating but relativistic phase, as
\begin{align}\label{RL}
R_{lat}&=\left(\frac{c E_0}{8 A}\right) t^{3/4}+\mathcal{O}(t^{5/4}).
\end{align}
We suggest that this explains the $R\propto t^{3/4}$ evolution of the apparent
size of SN 2012ap.

\section{Radio Emission}
The radio emission is powered by synchrotron losses suffered by accelerated electrons
in shock amplified magnetic fields. Assuming that a fraction $\epsilon_e$
of the total energy is used to accelerate electrons into a power law with
index $p=3$ extending from $\gamma_m$ to $\infty$ filling a fraction $f$ of the
volume, gives us the normalization of the electron distribution as,
\begin{equation}\label{N0}
 N_0=\frac{3 \epsilon_e E_0 \left(\gamma_m m_e c^2\right)^2}{2 \pi f R^3}
\end{equation}
Similar assumptions give the average magnetic field as
\begin{equation}\label{B}
 B=\sqrt{\frac{6 \epsilon_e E_0}{f R^3}}
\end{equation}
Once we have the evolution of $N_0$ and $B$, we can get the optically
thick and thin portions of the SSA spectrum as \citep{1979rpa..book.....R},
\begin{equation}
 F_\nu=\frac{\pi R^2}{D^2} \frac{c_5}{c_6} B^{-1/2} \left(\frac{\nu}{2c_1}\right)^{5/2},
\end{equation}
and
\begin{equation}
 F_\nu=\frac{4 \pi f R^3}{3 D^2} c_5 N_0 B^{(p+1)/2} \left(\frac{\nu}{2c_1}\right)^{-(p-1)/2},
\end{equation}
respectively, where $c_1$, $c_5$ and $c_6$ are constants \cite{1970ranp.book.....P} and
$D$ is the distance to the source.
In a mildly relativistic cases such as SN 2012ap, we may receive radiation from the entire
disk projected on the sky, hence $R$ is to be understood as $R_{lat}$. 
We then substitute for $N_0$ and $B$ from Equations \ref{N0} and \ref{B} and the leading
order expansion for the projected lateral radius from Equation \ref{RL}. The optically
thick and thin regimes meet at the turnover frequency of
\begin{equation}
 \nu_p=\frac{\left(4\ 2^{7/8} 3^{5/14} \left(\frac{A}{c}\right)^{23/56} c_1 E_0^{13/56} (c_6 \epsilon_e)^{2/7} \left(\frac{\epsilon_B}{f}\right)^{5/14} \left( \gamma_m m_e c^2\right)^{4/7}\right)}{\left(\pi^{2/7} t^{69/56}\right)}
\end{equation}
with a flux density of
\begin{equation}
 F_{\nu p}=\frac{4\ 2^{3/8} 3^{9/14} \pi^{2/7} c_5 \epsilon_e^{5/7}}{c_6^{2/7} D^2} \left(\frac{A}{c}\right)^{19/56}  \left(\frac{\epsilon_B}{f}\right)^{9/14} \left( \gamma_m m_e c^2\right)^{10/7}  \left(\frac{E_0}{t}\right)^{57/56} .
\end{equation}
Such an evolution of the SSA spectrum is qualitatively similar to that of type Ibc supernovae, in the
way the peak flux density decreases with time for both of them. But it is distinctly different
in the way that the peak flux density decreases in this case, while it is nearly constant
for supernovae in the nearly free expansion phase.

\section{Diagnostic Expressions}
In practice, the parameters of the explosion are not known a priori but the
peak flux density and frequency can be determined through observations. Therefore,
the above expressions must be inverted to solve for the explosion energy and circumstellar
density from the observed parameters.
%Total Energy
The expression for the explosion energy is most useful when put in terms of observable
quantities. We substitute the values of the constant c's and assume that
$\epsilon_e=\epsilon_B=0.1$ and $\gamma_m=100$ as is appropriate for a mildly relativistic
shock. This gives us
\begin{equation}
 E_0=2.35 \times 10^{45} \left(\frac{F_{\nu p}}{{\rm mJy}}\right)^{23/19}
 \left(\frac{\nu_p}{{\rm GHz}}\right)^{-1} \left(\frac{D}{{\rm Mpc}}\right)^{46/19} {\rm erg}.
\end{equation}

%Mass loss rate
The circumstellar density can be derived in a similar fashion. This gives us,
within the context of this relativistic blastwave model,
\begin{equation}
 \dot{M}=4.97 \times 10^{-10} \left(\frac{F_{\nu p}}{{\rm mJy}}\right)^{-13/19}
 \left(\frac{\nu_p}{{\rm GHz}}\right)^{3} \left(\frac{D}{{\rm Mpc}}\right)^{-26/19}
 \left(\frac{t}{{\rm day}}\right)^{3} \left(\frac{v_w}{10^3 {\rm km s^{-1}}}\right)
 {\rm M_\odot yr^{-1} .}
\end{equation}
We make a further
assumption that the wind speed was $v_w=1000 {\rm km s^{-1}}$ as is appropriate
for a Wolf Rayet progenitor. Using the data from age 18 days, we estimate the
pre-explosion mass loss rate as $(3.5\pm 0.2)\times10^{-6} {\rm M_\odot yr^{-1}}$.

The above model is strictly true for highly relativistic blastwaves, while the observed
explosion is only mildly relativistic. We therefore, also consider the Newtonian
limit to see if it gives systematically different results.
The entire calculation can be redone in the context of the Newtonian Sedov-Taylor-like
blastwave. The final result for the explosion energy remains the same, indicating that
the calorimetric estimate of the explosion energy is robust. The expression for
the circumstellar density changes to,
\begin{equation}
 \dot{M}=1.42 \times 10^{-8} \left(\frac{F_{\nu p}}{{\rm mJy}}\right)^{-4/19}
 \left(\frac{\nu_p}{{\rm GHz}}\right)^{2} \left(\frac{D}{{\rm Mpc}}\right)^{-8/19}
 \left(\frac{t}{{\rm day}}\right)^{2} \left(\frac{v_w}{10^3 {\rm km s^{-1}}}\right)
 {\rm M_\odot yr^{-1} .}
\end{equation}
This gives us an estimate of $(4.9\pm 0.3)\times10^{-5} {\rm M_\odot yr^{-1}}$.
The real mass loss rate is possibly in between these two systematically different
estimates. Upper limits on the mass loss rate from \citet{2014arXiv1402.6344M}
rule out this higher mass loss rate and are consistent with the results from the
relativistic model.

%% The reference list follows the main body and any appendices.
%% Use LaTeX's thebibliography environment to mark up your reference list.
%% Note \begin{thebibliography} is followed by an empty set of
%% curly braces.  If you forget this, LaTeX will generate the error
%% "Perhaps a missing \item?".
%%
%% thebibliography produces citations in the text using \bibitem-\cite
%% cross-referencing. Each reference is preceded by a
%% \bibitem command that defines in curly braces the KEY that corresponds
%% to the KEY in the \cite commands (see the first section above).
%% Make sure that you provide a unique KEY for every \bibitem or else the
%% paper will not LaTeX. The square brackets should contain
%% the citation text that LaTeX will insert in
%% place of the \cite commands.

%% We have used macros to produce journal name abbreviations.
%% AASTeX provides a number of these for the more frequently-cited journals.
%% See the Author Guide for a list of them.

%% Note that the style of the \bibitem labels (in []) is slightly
%% different from previous examples.  The natbib system solves a host
%% of citation expression problems, but it is necessary to clearly
%% delimit the year from the author name used in the citation.
%% See the natbib documentation for more details and options.

%\clearpage

\bibliographystyle{apj}
\bibliography{master}

%\clearpage

%% Use the figure environment and \plotone or \plottwo to include
%% figures and captions in your electronic submission.
%% To embed the sample graphics in
%% the file, uncomment the \plotone, \plottwo, and
%% \includegraphics commands
%%
%% If you need a layout that cannot be achieved with \plotone or
%% \plottwo, you can invoke the graphicx package directly with the
%% \includegraphics command or use \plotfiddle. For more information,
%% please see the tutorial on "Using Electronic Art with AASTeX" in the
%% documentation section at the AASTeX Web site,
%% http://www.journals.uchicago.edu/AAS/AASTeX.
%%
%% The examples below also include sample markup for submission of
%% supplemental electronic materials. As always, be sure to check
%% the instructions to authors for the journal you are submitting to
%% for specific submissions guidelines as they vary from
%% journal to journal.

%% This example uses \plotone to include an EPS file scaled to
%% 80% of its natural size with \epsscale. Its caption
%% has been written to indicate that additional figure parts will be
%% available in the electronic journal.

\end{document}